\documentclass[twocolumn,showpacs,prl,superscriptaddress]{revtex4}

\usepackage{graphicx}  
\usepackage{dcolumn}   
\usepackage{bm}        

\usepackage{amsmath}
\usepackage{amssymb}
\usepackage{array}
\usepackage[english]{babel}

\newcommand{\efield}{\bm{\mathcal E}} 
\newcommand{\bhat}[1]{\hat{\bm #1}}
\newcommand{\acphase}{\hat{\bm \Phi}}
\newcommand{\hamilt}{\hat H}
\newcommand{\iop}{\hat{\bm 1}}
\newcommand{\phiac}{\Phi_{AC}}
\newcommand{\AC}{A\mbox{-}C }

\newcommand{\nonum}{} 

\begin{document}

\preprint{PRL}

\title{Single atom-scale diamond defect allows large Aharonov-Casher phase}

\author{D. Maclaurin}
\affiliation{School of Physics, The University of Melbourne, Parkville,
3010, Australia}
\affiliation{Centre for Quantum Computer Technology, School of Physics, The University of Melbourne, Parkville 3010, Australia}
\author{A.D. Greentree}
\affiliation{School of Physics, The University of Melbourne, Parkville,
3010, Australia}
\author{J.H. Cole}
\affiliation{Institut f\"ur Theoretische Festk\"orperphysik und DFG-Center for Functional Nanostructures (CFN),\\ Universit\"at Karlsruhe, 76128 Karlsruhe, Germany}
\author{L.C.L. Hollenberg}
\affiliation{School of Physics, The University of Melbourne, Parkville,
3010, Australia}
\affiliation{Centre for Quantum Computer Technology, School of Physics, The University of Melbourne, Parkville 3010, Australia}
\author{A.M. Martin}
\affiliation{School of Physics, The University of Melbourne, Parkville,
3010, Australia}

\date{\today}

\begin{abstract}
We propose an experiment that would produce and measure a large Aharonov-Casher (A-C) phase in a solid-state system under macroscopic motion. A diamond crystal is mounted on a spinning disk in the presence of a uniform electric field. Internal magnetic states of a single NV defect, replacing interferometer trajectories, are coherently controlled by microwave pulses. The A-C phase shift is manifested as a relative phase, of up to 17 radians, between
components of a superposition of magnetic substates, which is two orders of magnitude larger than that measured in any other atom-scale quantum system.
\end{abstract}

\pacs{03.65.Vf, 76.30.Mi, 42.50.Dv, 03.65.Yz}

\maketitle


One of the fundamental predictions of quantum mechanics is the existence of topological phases. A famous case still relatively unexplored is the Aharonov-Casher (A-C) effect: if a particle with a magnetic moment is taken about a line of charge, it acquires a phase which is independent of its velocity and only topologically dependent on its path \cite{AC}. The effect is important for spin currents in mesoscopic rings \cite{Balatsky,Bergsten}, vortices in superconductors \cite{Wees,Elion}, and it has potential applications in topological quantum computing \cite{Reuter,Kitaev}. The \AC phase, a relativistic effect, is much harder to produce than the Aharonov-Bohm effect \cite{AB} and other better-studied topological phases. Experiments using atom beams \cite{Sangster, Gorlitz, Zeiske} and neutron interferometers \cite{Cimmino} have successfully measured the A\mbox{-}C effect but only small phases have been produced, of order 0.15 rad or less. In contrast to these atom-scale systems, large A-C phases can arise through collective effects in mesoscopic rings \cite{Bergsten}. Here, we propose a method to generate and measure a large \AC phase through quantum control of a single atomic-scale system: the negatively charged diamond nitrogen-vacancy (NV) center.

\begin{figure}
\centering
\includegraphics[width=8.5cm]{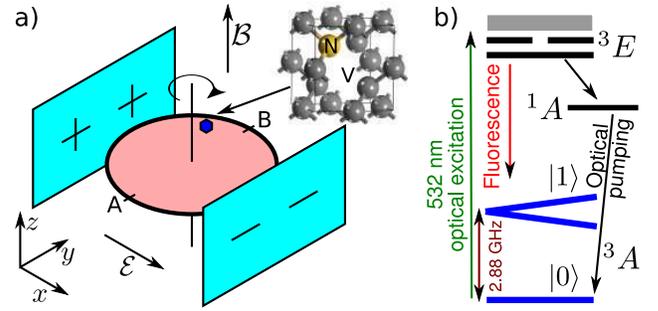}
\caption{Proposed experimental setup. (a) Geometry of experiment; a diamond crystal on a spinning disk between two charged plates with a uniform, static magnetic field in the $z$-direction. (b) Energy level diagram of the NV center. \label{one}}
\end{figure}

The NV center is a naturally occurring defect in diamond which has a spin triplet electronic ground state with excellent coherence properties, making it a widely considered candidate for a quantum bit \cite{Greentree,Dutt}. In our proposed experiment, an NV center, coherently controlled by microwave pulses, acquires a relative \AC phase between its magnetic sublevels as it moves between two charged plates (Fig. \ref{one}(a)). The experiment is made possible by recent developments in coherent control of NV centers, which have shown that an NV center can act as a highly sensitive magnetometer, detecting fields down to nTHz$^{-1/2}$ sensitivity, at room temperature \cite{Harvard_exp,Stuttgart_exp,Taylor,Nature_mat,Degen,Decoherence}. The experiment would be a powerful test of the \AC effect, generating an accurately measurable \AC phase two orders of magnitude larger than those of previous experiments, and could open up a new field of applications for coherently controlled solid-state quantum systems.

\textit{\AC phase.}---Aharonov and Casher considered a neutral, non-relativistic, spin-$\frac{1}{2}$ particle, with magnetic dipole moment $\mu \bhat{\sigma}$, moving in a chargeless region of nonvanishing electric field $\efield(\bm x)$. The Hamiltonian for such a particle can be derived from the Dirac Lagrangian in the low-energy limit:
\begin{equation}
\hamilt = \frac{1}{2m}\left [\bhat{p} - \frac{\mu}{c^2} \bhat \sigma \times \efield(\bhat x)\right ]^2 - \frac{\mu^2\efield(\bhat x)^2}{2mc^4}. \nonum \label{Ham_1}
\end{equation}
The \AC effect is usually treated in 2+1 dimensions. Some authors use an explicitly two-dimensional wave equation \cite{McKellar}. Others effectively reduce the Hamiltonian to two dimensions by requiring the wavefunction to be independent of the $z$-coordinate, and setting $\mathcal E_z = 0$ and $\partial_z \efield = 0$ \cite{Hagen}. In either approach, if $|\psi_0\rangle$ is some solution to the Schr\"{o}dinger equation in the absence of an electric field, then
\begin{eqnarray}
|\psi\rangle &=& e^{-i\acphase(\hat x,\hat y)}|\psi_0\rangle, \nonum \\
\text{ where }\quad \acphase(x,y) &=& -\frac{\mu}{\hbar c^2}\int_\gamma \bhat \sigma \times \efield(x',y')\cdot d\bm x' \label{AC_1}
\end{eqnarray}
solves the Schr\"{o}dinger equation in the presence of the electric field. The path of integration $\gamma$ can be any path in the $x$-$y$ plane that shares the same topology as the path of the particle. $|\psi\rangle$ differs from $|\psi_0\rangle$ only by an SU(2) phase $\acphase (x,y)$ which is the \AC phase.

We briefly outline an alternative derivation of the \AC effect, which holds for a localized particle of arbitrary spin in three dimensions. For particles of arbitrary spin $S$, the Dirac equation generalizes to the Bargmann-Wigner (B-W) equation \cite{McKellar}. Taking the B-W equation in the low-energy limit gives the Hamiltonian
\begin{eqnarray}
\hamilt &=& \frac{1}{2m}[\bhat p - \frac{\mu}{c^2} \bhat S \times \efield(\bhat x)]^2 + \hat V(\bhat x)\nonumber \\
        &+& \frac{\mu^2\efield(\bhat x)^2}{2mc^4}-\frac{1}{2mc^4}[\mu \bhat S \times \efield(\bhat x)]^2, \label{BW_hamilt}
\end{eqnarray}
Where $\bhat S$ is the spin operator of the particle, $\hat V$ is a potential which may depend on spin, and $\mu \bhat{S}$ is the magnetic moment of the particle. We can ignore the last two terms of Eq. (\ref{BW_hamilt}) since, due to the pulse sequence discussed below, they will not contribute to the measured signal.

Let $|\psi_0 ; t \rangle$ be a solution to the Schr\"{o}dinger equation
with $\mathcal E = 0$, that describes a highly localized particle following a
trajectory $\bm r (t)$. A solution to the Schr\"{o}dinger equation with $\mathcal E \neq 0$ can then be written as
\begin{eqnarray}
            |\psi ; t \rangle &=&\hat A(t) \hat U(\bm r(t)) |\psi _0; t \rangle, \nonum \\
\text{where} \qquad \hat A(t) &=& 1 + \frac{i\mu}{\hbar c^2}\bhat S\times \efield(\bm r(t)) \cdot [\bhat x - \bm r(t)], \nonum
\end{eqnarray}
And $\hat U$ must satisfy an integral equation,
\begin{equation}
\hat U(\bm r(t)) = 1 + \frac{i\mu}{\hbar  c^2}\int_0^{t}[ \bhat S\times \efield (\bm r(t')) \cdot \dot{\bm r}]\hat U(\bm r(t')) dt'
\label{int_exp}
\end{equation}
whose solution can be expressed as a Dyson series. We have assumed that $[\hat U(\bm r(t)),\hat V] = 0 $, and that the particle is sufficiently localized that $\bhat x|\psi_0 ; t \rangle \approx \bm r (t) |\psi_0 ; t \rangle$ and, hence, $\hat A|\psi_0\rangle \approx |\psi_0\rangle$.

The operator $\hat U(\bm x)$ is an SU(2) phase factor, similar to the \AC phase factor $\text{exp}[-i\acphase(x,y)]$ of Eq. (\ref{AC_1}). However, whereas $\acphase(x,y)$ only depends on the path's topology, $\hat U(\bm x)$ depends, in general, on the details of the path $\bm r(t)$ of the particle. Small movements of the particle cause precession of its spin about an axis perpendicular to both the electric field and the direction of motion. The path-dependence of $\hat U$ arises from the non-commutativity of rotations about different axes: as Anandan notes \cite{Anandan}, we can view $\bhat S\times \efield$ as a non-Abelian gauge field.

The full SU(2) phase can be explored by tilting the plane of the spinning disk about the $y$-axis. However, the non-Abelian effects will be small, as they are of second order in the Dyson series expansion of (\ref{int_exp}). If we restrict the particle's motion to the $x$-$y$ plane, and set $\mathcal E_z = 0$ and $\partial_z \efield = 0$ as before, we recover the usual, path-independent, \AC effect: $\hat U(\bhat x) = e^{-i\acphase}$.

\textit{\AC phase in diamond.}---The NV electronic energy level structure is shown in Fig. \ref{one}(b). We are interested in the $^3A$ ground state, which is connected to the
excited states by optical excitations. Initialization of the NV center
is performed by optical pumping into $|0\rangle$, followed by Rabi
pulses to achieve some coherent superposition of the spin sublevels
$|0\rangle$ and $|1\rangle$ \cite{Jelezko}. The relative populations of the spin sublevels can be observed by optical excitation and a measurement of fluorescence.

Ignoring crystal asymmetries and interactions with nuclear spin, the
spin Hamiltonian $\hamilt_{s}$ for the $^3$A ground state can be written as
\begin{equation}
\hamilt_s = D(\hat S_z -\frac{1}{3}\bhat S^2) + g\mu_B\bm B\cdot
\bhat S, \nonum
\end{equation}
where $D = 2.88$ GHz is the zero-field splitting parameter, $\bm B$ is
the external magnetic field, and $g\approx 2$ is the gyromagnetic ratio of
the NV center. Note the NV center's negative charge does not affect the spin Hamiltonian and hence is decoupled from the A-C dynamics.

To treat the \AC effect for the NV center, we introduce a Hilbert space that
is spanned by the position eigenstates of the NV center. A general state of
the NV center can then be written as $|\psi\rangle = |\psi_S\rangle \otimes |\psi_R \rangle$ where $|\psi_S\rangle$ is a superposition of $|0\rangle$, $|1\rangle$
and $|\text{-}1\rangle$, and $|\psi_R \rangle$ gives the position of the NV center. The total Hamiltonian of $^3A$ state is then
\begin{equation}
\hat H = \hat H_s \otimes \iop_R + \frac{1}{2m}[\iop_S \otimes \hat{\bm p}+
\frac{g\mu_B}{c^2}(\bhat S\times \efield) \otimes \iop_R ]^2 + \iop_S \otimes
\hat V, \nonum
\end{equation}
where $\hat{\bm p}$ is the NV center-of-mass momentum, $m$ is
the NV center's effective mass, and $\hat V$ is a potential which acts only
on the position space.

Now consider an NV center moving in the $x$-$y$ plane, in the presence
of an electric field $\efield(\bm x)$, with $\mathcal E_z = 0$ and $\partial_z \efield = 0$, and a
magnetic field, of magnitude $B$, in the $z$-direction. Let $|\bm r(t)\rangle$ be
a solution to the Schr\"{o}dinger equation for the position Hilbert space, such that $|\bm
r(t)\rangle$ describes the motion of the NV center along some
trajectory, $\bm r(t)$. If the NV center is initialized to
\begin{equation}
|\psi;t=0\rangle = (c_0|0\rangle +  c_1 |1\rangle) \otimes |\bm r(t=0)\rangle
\end{equation}
where the $c_i$ are complex coefficients, then after time $t$ the
state of the NV center will be
\begin{eqnarray}
|\psi;t\rangle &=& \big [c_0|0\rangle +
e^{-i\phiac(\bm r(t))}e^{-it(D+gB\mu_B)/\hbar} c_1            |1\rangle \big ]  \nonumber\\
&\otimes& |\bm r(t)\rangle, \label{evolution}
\text{where} \quad \phiac(\bm x) = \frac{g\mu_B}{\hbar c^2}\int_{\gamma}
\hat{\bm k} \times \bm E(\bm x')\cdot  d\bm x'.
\end{eqnarray}
It is this $\phiac(\bm x)$, the \AC phase in diamond, which our proposed experiment would measure.

We briefly pause to consider the topological nature of the AC effect.
Boyer \cite{Boyer} and Casella \cite{Casella}, among others, have
contested the appropriateness of the term `topological'. In particular, Casella noted that the \AC effect can be detected without a topological defect and proposed the geometry that we use here: a
particle in a superposition of states passes between two charged
plates. (The \AC phase in this geometry is often termed the `Casella
phase'.) G\"{o}rlitz \textit{et al.} \cite{Gorlitz} make a strong case for observing the \AC phase using this geometry, in which the superposed substates are treated as arms of an interferometer. The characteristic nondispersiveness and path-independence of topological phases remain in any case.

\textit{Experimental proposal.}---We envisage an experimental setup as
illustrated in Fig. \ref{one}(a). A diamond crystal is embedded at the edge
of a disk of radius $r$ that lies in the $x$-$y$ plane and spins, with
frequency $f$, about the $z$-axis. The crystal is oriented so that the NV of interest is aligned parallel to the $z$-axis. Two charged plates, connected to a constant voltage source, provide a uniform
static electric field $\mathcal E$ in the $x$-direction. A pair of Helmholtz coils provides a uniform static magnetic field $B$, on the order of Gauss, along the $z$-direction, which lifts the $m=\pm 1$ degeneracy \cite{Harvard_exp}. The experiment consists of initializing a single NV center into a coherent superposition of spin sublevels and observing the relative phase that accumulates between the sublevels due the \AC effect. As the disk rotates, the NV spin states evolve according to (\ref{evolution}). Note that the rate of \AC phase accumulation varies sinusoidally in time. Positive (negative) \AC phase accrues as the diamond moves in the positive (negative) $y$-direction.

\begin{figure}
\centering
\includegraphics[width=8.5cm]{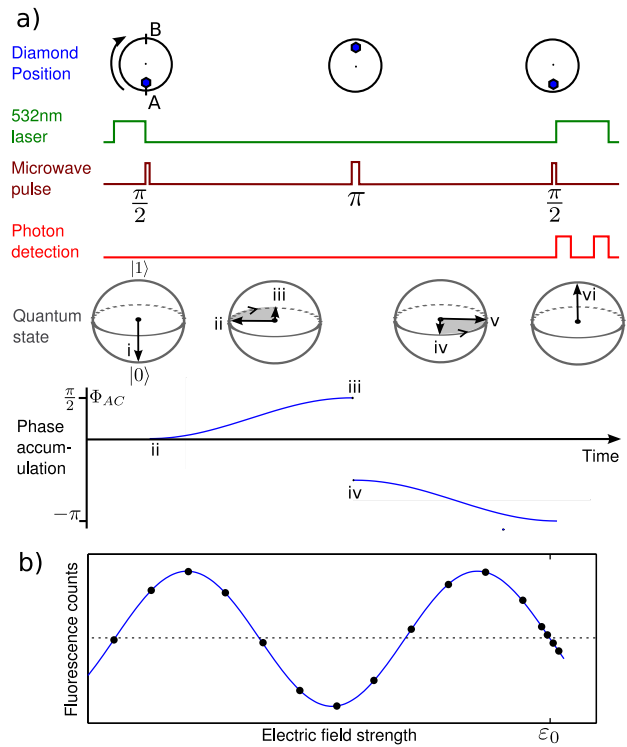}
\caption{(a) Pulse scheme (pulse times not to scale) and rotating-frame Bloch sphere state evolution for a single rotation of the disk, with $\pi$ rad of accumulated \AC phase. Runs with multiple rotations would have further $\pi$ pulses whenever the diamond passes points A and B. (b) Example fluorescence signal as a function of electric field strength, where the maximum \AC phase is 10 rad. $y$-axis units are arbitrary. Dots represent example experimental data points. \label{two}}
\end{figure}

The NV center is initialized, controlled and measured using a 532nm laser, a microwave generator, and a photodetector, which each focus on particular points of the diamond's path as shown in Fig. \ref{two}(a). The laser is
used for optical pumping and readout and the photodetector measures fluorescence. The microwave generator drives two coils, located close to the spinning disk, at points A and B, which perform controlled Rabi rotations of the NV center's spin sublevels. Alternatively, Rabi oscillations could be induced optically \cite{Santori}.

The pulse sequence and the corresponding quantum state evolution are shown schematically in Fig. \ref{two}(a). This is
essentially a spin echo experiment: $\pi$ pulses are used to remove
any net precession due to static magnetic fields, while rectifying
the sinusoidal \AC phase accumulation. The disk is first set
spinning. As the diamond passes point A, the NV center is optically
pumped into the $m=0$ state $|0\rangle$ using the $532$nm laser (Bloch sphere i of Fig \ref{two}(a)).
Next, a $\pi/2$ Rabi pulse creates a coherent superposition of
$|0\rangle$ and $|1\rangle$ (ii). As the disk continues to rotate, a $\pi$
Rabi pulse is applied every time the diamond passes points A or B,
which inverts the phase between the $|0\rangle$ and $|1\rangle$
sublevels (iii,iv). After the diamond has completed $n$ full rotations (v), a second $\pi/2$
pulse is applied, which converts the relative phase
acquired during rotation into a population difference between the sublevels (vi). A pulse
from the $532$nm laser then excites the NV center and a fluorescence
measurement is taken. Note that we can ignore the motion of the diamond for the duration of each of the pulses ($\sim$0.5 $\mu$s  or less): given reasonable figures for the disk's rotation frequency and radius (see below), the diamond is stationary to within a few tens of microns. The total phase accumulated $\Phi$ is given by
\begin{equation}
\Phi = \frac{4}{\hbar c^2} g\mu_B r \mathcal En. \nonum
\end{equation}

An ensemble of measurements, made over time, will not be perfectly coherent. For NV centers, the dominant decoherence process is dephasing (divergence in the phase evolution of different members of the ensemble) rather than spin relaxation. Dephasing manifests itself as an exponential decay of the fluorescence signal, towards a value corresponding to equal populations of $|0\rangle$ and $|1\rangle$, as the evolution time of the experiment is increased.

The fastest dephasing, due to inhomogeneous magnetic fields, gives a so-called $T_2^*$ dephasing time on the order of $\mu$s \cite{Childress}. However, these magnetic fields fluctuate slowly compared with other time scales, which means that the sequence of $\pi$ pulses eliminates any net effect \cite{Harvard_exp}. The next fastest dephasing is the $T_2$ homogeneous broadening due fast-fluctuating magnetic fields caused by the nuclear spin of $^{13}C$ atoms. $T_2$ times of up to 1.8 ms have been recorded in recent experiments \cite{Nature_mat}.

If the pulse sequence is conducted for $\mathcal E$ ranging from zero to $\mathcal E_0$ while keeping the number of rotations per run $n$ and the evolution time $t_r$ fixed, then the fluorescence count, plotted as a function of $\mathcal E$, will look like Fig. \ref{two}(b). To minimize error in measuring the maximum \AC phase from the fluorescence signal, the gradient of the curve should be a maximum at $\mathcal E_0$. This can be achieved by appropriately shifting the phase of the final $\pi/2$ microwave pulse. In the example shown, the pulse's phase lags by 2.15 rad behind the earlier pulses. To measure a very small phase ($\phiac \ll 1$ rad), the lag should be $\pi/2$.

Van Oort has shown that, in addition to the well-known excited state Stark shift \cite{Redman}, the NV ground state experiences a linear Stark
shift \cite{Van_Oort}, with Hamiltonian
\begin{equation}
H_{\text{Stark}} = -\hbar \mathcal E R_{2E} \left (e^{-3i\theta}|\text{-}1\rangle\langle 1|+e^{3i\theta}|1\rangle\langle \text{-}1|\right ),
\label{H_stark}
\end{equation}
where the constant $R_{2E}$ is approximately 20 HzV$^{-1}$cm and $\theta$ is the angle between $\efield$ and one of the crystal's three planes of symmetry containing the N-V axis. The energy eigenstates then rotate at triple the frequency of the diamond's rotation. By the adiabatic theorem, provided the diamond's rotation frequency is much smaller than the energy separation between the $m=\pm 1$ states, the Stark shift will simply shift the energy of these states by a constant amount, $\hbar^2 R_{2E}^2\mathcal E^2/(B_z g \mu_B)$. The $\pi$-pulses will therefore eliminate any net effect of the ground state Stark shift.

We envisage realistic experimental parameters $f=4$ kHz, $r=1$ cm, $\mathcal E=30$ kV/mm \cite{Klein}, and $T_2 = 1.8$ ms. These parameters would give a total \AC phase of $\sim 17$ rad, which is two orders of magnitude larger than previous neutron and atom beam experiments \cite{Gorlitz,Zeiske}. Experiments measuring the resistance of mesoscopic rings have demonstrated A\mbox{-}C-like phases of this order \cite{Bergsten}, though in these cases the phase comes from precession about a continuously changing axis, as the electric field is perpendicular to the plane of motion of the magnetic dipole, and the Rashba spin-orbit coupling arises from a many-body interaction.

Two intrinsic sources of error are shot noise, due to the finite
fraction of photons detected, and spin projection noise, due
to the probabilistic nature of quantum measurements. These both follow Poisson statistics, yielding a sensitivity $\eta$ in the measurement of $\Phi_{AC}$,
\begin{equation}
\eta=\frac{\Delta \Phi_{\text{AC}}}{\sqrt{T}} \approx \frac{\sqrt{2}}{C \sqrt{T_2}}.
\label{error}
\end{equation}
$C$ ($\approx 0.05$ for typical experiments) is a factor which takes into account collection efficiency and the nonzero fluorescence of $|1\rangle$ \cite{Taylor} and $T$ is the combined time of the individual runs. We have taken the duration $t_r$ of each run to be $T_2$, to optimize the relative uncertainty $\Delta \Phi_{\text{AC}}/\Phi_{\text{AC}}$ given the signal's exponential decay in $t_r/T_2$. 

A single NV center could measure an \AC phase to within a radian over
100 hours ($\eta \approx $ 700 rad Hz$^{-1/2}$). Alternatively, a spatial ensemble of N individual centers would allow a high-precision measurement, improving the sensitivity by $\sqrt{N}$. NV densities as high as 10$^{11}$mm$^{-3}$ could be used without admitting spin-spin decoherence \cite{Taylor} which, given a crystal size of (1 mm)$^3$, would give very high sensitivity: $\eta \approx $ 2 mrad Hz$^{-1/2}$.

We have shown that an A-C phase shift in diamond can be observed using current experimental techniques. The phase shift can be very large and can be measured to great accuracy. Moreover, as a solid state system whose path can be precisely controlled, the NV center can be used to
probe deeper aspects of topological phases, such as the non-Abelian dynamics of the \AC phase in three dimensions.

\begin{acknowledgments}
We thank A. Klein and B. McKellar for valuable conversations. This work was supported by the Australian Research Council (Projects DP0880466 and DP0770715) and the Alexander von Humbolt Foundation.
\end{acknowledgments}


\end{document}